\documentclass{article}
\usepackage[english]{babel}
\usepackage{amsmath,amsfonts,amsthm,amssymb,amscd}

\renewcommand{\Re}{\mathop{\rm Re}\nolimits}

\newcommand{\e}{\varepsilon}

\newcommand{\R}{{\mathbb R}}

\newcommand{\pP}{{\mathbb P}}
\newcommand{\E}{{\mathbb E}}

\newcommand{\N}{{\mathbb N}}
\newcommand{\la}{\lambda}

\newcommand{\BBB}{{\boldsymbol{B}}}
\newcommand{\HHH}{{\boldsymbol{\mathrm{H}}}}

\newcommand{\mmm}{{\mathfrak{m}}}

\newcommand{\BB}{{\cal B}}
\newcommand{\DD}{{\cal D}}

\newcommand{\FF}{{\cal F}}
\newcommand{\GG}{{\cal G}}
\newcommand{\HH}{{\cal H}}

\newcommand{\LL}{{\cal L}}

\newcommand{\NN}{{\cal N}}

\newcommand{\PP}{{\cal P}}
\newcommand{\RR}{{\cal R}}

\newcommand{\dd}{{\textup d}}

\newcommand{\PPPP}{{\mathfrak P}}

\newcommand{\BBBBB}{{\mathcal B}}

\def\S{\mathhexbox278}

\theoremstyle{plain}
\newtheorem{theorem}{Theorem}[section]
\newtheorem{lemma}[theorem]{Lemma}
\newtheorem{proposition}[theorem]{Proposition}

\newtheorem{definition}[theorem]{Definition}
\newtheorem{condition}[theorem]{Condition}
\theoremstyle{remark}
\newtheorem{remark}[theorem]{Remark}

\newcommand{\de}{\delta}
\numberwithin{equation}{section}
\begin{document}
\author{Vahagn Nersesyan}
\date{}
\title{Polynomial mixing for the complex Ginzburg--Landau equation perturbed by a random force at random times}
\date{}
 \maketitle
\begin{center}
 Laboratoire de Math\'ematiques,
Universit\'e de Paris-Sud XI\\ B\^atiment 425, 91405 Orsay Cedex,
France\\ E-mail: Vahagn.Nersesyan@math.u-psud.fr
\end{center}
{\small\textbf{Abstract.} In this paper we study the problem of
ergodicity for the complex Ginzburg--Landau (CGL) equation perturbed
by an unbounded random kick-force. Randomness is introduced both
through the kicks and through the times between the kicks. We show
that the Markov process associated with the equation in question
possesses a unique stationary distribution and satisfies a
property of polynomial mixing.}\\\\
\section{Introduction}\label{intro}
We consider the CGL equation perturbed by a random kick-force on a
domain $D\Subset\R^n,$ $n\le 4$ with $\partial D\in C^2:$
\begin{align}
\dot{u}-\nu\Delta u+i\beta|u|^2u&=\eta(t,x),\,\,\, x\in D,
\label{E:eq}\\
 u\arrowvert_{\partial D}&=0,\label{E:lc}\\
u(0,x)&=u_0(x),\label{E:ic}
\end{align}
where $u=u(t,x)$ and $\nu,\beta>0.$ We assume that $\eta(t,x)$ is a
random process of the form
\begin{equation}\label{E:pr}
\eta(t,x)=\sum_{k=1}^{\infty}\eta_k(x)\delta(t-\tau_k),
 \end{equation}
where $\delta(t)$ is the Dirac measure, $\eta_k$ are independent
identically distributed (i.i.d.) random  variables with range in the
space $H:=H_0^1(D)$, and the waiting times
$t_k=\tau_{k}-\tau_{k-1},$ $k\ge2$ and  $t_1=\tau_1$ are i.i.d.
random variables exponentially distributed with parameter $\lambda$.
Moreover, we assume that the sequences $\eta_k$, $t_k$ are
independent.


Suppose that $\{g_k\}_{k=1}^{\infty}$ is an orthonormal basis in
$H$. The main result of the present paper is Theorem \ref{T:him},
which states that, if the low of $\eta_k$ is non-degenerate on the
space spanned by $\{g_k\}_{k=1}^N$ for sufficiently large $N$, then
there is a unique stationary measure for the continuous time Markov
process associated with (\ref{E:eq}), (\ref{E:lc}), (\ref{E:pr}).
Moreover, any solution of the problem polynomially converges to the
stationary measure in the dual Lipschitz norm.

 Many authors have studied similar problems for various PDE's with different random
perturbations (e.g., see \cite{KS14,BKL,KS15,KSH,MY,KPS2,Matt,K12,Shi04}
for discrete forcing and
\cite{FL,EMS,BKL2,KSC,EH,Hai,Od,ShiII,DOd}  for white noise).
Several ideas of this article are taken from \cite{KSH,KPS2,Shi04}.


The problem of ergodicity for randomly forced Ginzburg--Landau
equation was studied in the  following articles. In \cite{Hai},
Hairer considered a real  Ginzburg--Landau equation on
multidimensional torus. Odasso \cite{Od} studied a class of CGL
equations with strong nonlinear dissipation. In both of these works
the property of exponential mixing is established. In \cite{ShiII},
Shirikyan used a sufficient condition for ergodicity of Markov
processes to show uniqueness and mixing for a class of CGL equations
with linear dispersion. Finally, in \cite{DOd}, Debussche and Odasso
proved the polynomial mixing property for a damped 1D Schr\"odinger
equation.

The main novelty of the present paper is the condition over the
waiting times. Note that the restriction of the solution at times $\tau_k$ looks like the random dynamical systems considered by  Kuksin, Shirikyan  \cite{K12},
 \cite{KS14}, 
 \cite{Shi04}
  and Masmoudi, Young~\cite{MY}
:
\begin{equation}\label{E:embd}
u_{\tau_k}=S_{t_k}(u_{\tau_{k-1}})+\eta_k,
\end{equation}
 but there are some essential differences.
As the waiting times can be arbitrarily small, during any time interval the system can receive any number of kicks.  This changes the dynamics of the associated process, for example:
\begin{enumerate}
\item[$\bullet$] The distance  between two trajectories having close initial data can be arbitrary large at any time $t>0$.
\item[$\bullet$]The phase space of the problem is not bounded even in the case of bounded kicks.
\end{enumerate}

Let us give in a few words the ideas of the proof of Theorem
\ref{T:him}. An important tool for the proof of the result is the
Foia\c{s}--Prodi type estimate. This kind of estimates are often
used to prove ergodic properties of PDE's. Suppose that there are
two sequences of kicks $\zeta_k$ and  $\zeta_k'$, having equal high
Fourier modes for $k\ge l$, such that the solutions of corresponding
problems have equal low Fourier  modes at kicking times $\tau_k$,
$k\ge l$  (see Lemma \ref{L:P} for the exact formulation). Let
$\NN_t$ be the number of kicks before time $t$, i.e. $\NN_t=\max\{k:
\tau_k\le t\}$. Then, by  Foia\c{s}--Prodi Lemma,   we have the
following estimate for the distence between solutions at time $t$,
if $t\ge\tau_{l}$:
\begin{equation}\label{E:Ha12}
\|u_t-u_t'\|_1\le e^{-C(\NN_t-l)}\big(\prod_{i=l+1}^{\NN_t}t_i\big)^{-\frac{1}{2}}e^{\varphi}\|u_{\tau_l}-u_{\tau_l}'\|_1,
\end{equation}
where $\|\cdot\|_1 $ stands for the norm in $H$, $u_t$ and $u_t'$ are solutions corresponding to the sequences $\zeta_k$ and  $\zeta_k'$ respectively,  $\varphi$ is a polynomial function of $\{\|u_{\tau_i}\|_1\}_{i=l}^{\NN_t}$ and $\{\|u_{\tau_i}'\|_1\}_{i=l}^{\NN_t}$ and $C>0$ is a large constant.
Following the ideas from \cite{Shi04}, we construct two sequences  $\zeta_k$ and  $\zeta_k'$ of i.i.d. random variables in $H$ distributed as $\eta_1$ such that the conditions of  Foia\c{s}--Prodi Lemma are satisfied for a random integer $\ell\ge1$. Moreover, using the low of large numbers and some martingale inequalities, we show that $\ell$ can be choosen in a such way that the following properties also hold:
\begin{enumerate}
\item[(i)]$\big(\prod_{i=\ell+1}^{\NN_t}t_i\big)^{-\frac{1}{2}}e^{\varphi}\le e^{(\NN_t-\ell)}$, if $\NN_t\ge \ell+1,$
\item[(ii)]$\|u_{\tau_\ell}\|_1+ \|u_{\tau_\ell}'\|_1\le1$,
\item[(iii)]$\E \ell^p\le C_p$.
\end{enumerate}
As we show in Section \ref{MR}, properties (i)-(iii) and (\ref{E:Ha12})
 imply the polynomial mixing property.

 The random variables $\zeta_k,\zeta_k'$ and $\ell$ are constructed in Proposition \ref{P:A}. 
 In  Section \ref{MR}, we show properly  how  Theorem \ref{T:him} is derived from Proposition \ref{P:A}. The proof of  Proposition \ref{P:A} is  carried out in Sections \ref{CO} and \ref{PP}.

Note that an exponential estimate for the random variable $\ell$ in
(iii) implies immediately the exponential mixing property for the
system. Finally, using (i)-(iii), one can show that the embedded
Markov chain $u_{\tau_k}$ also satisfies a property of polynomial
mixing. The stationary measure of the original process and that of
embedded chain are connected with the Khasminskii relation (see
Section \ref{MR}).

\textbf{Acknowledgments.} The author thanks A. Shirikyan for
attracting his attention to this problem and for helpful discussions
and encouragements. A part of this paper was written when the author
was visiting the De Giorgi Center (Pisa); he thanks the Center for
hospitality.
\\
\\

\textbf{Notation}
\\\\
Let $D\subset \R^n$ be a  bounded domain with smooth boundary and
let $\{g_j\}_{j\in\N}$ be an orthonormal basis in $H$.  Let $H_N$ be
the vector span of $\{g_1,...,g_N\}$ and  $H_N^{\bot}$ be its
orthogonal complement in $H$. We denote by $P_N$ and $Q_N$ the
orthogonal projections onto $H_N$ and $H_N^{\bot}$ in $H$. Denote by
$\{e_j\}_{j\in \N}$ the set of normalized eigenfunctions of the
Dirichlet Laplacian with eigenvalues  $\{\alpha_j\}_{j\in \N}$ and
denote by $Q_N'$ the orthogonal projection onto the closure of the
vector span of $\{e_N,e_{N+1},...\}$ in $L^2(D)$.
\\ Let  $H^s(D)$, $s\ge 0$ be the Sobolev space of order $s$. We
denote by $\|u\|_1=\|\nabla u\|, \|u\|_2=\|\Delta u\|$ the norms in
the spaces $H_0^1(D)$ and $H_0^1(D)\cap H^2(D)$ respectively, where
$\|\cdot\|$ stands for the norm in $L^2(D).$ For a Banach space $X$,
we shall use the following notation.
\\ $\BBBBB(X)$ is the $\sigma$-algebra of Borel subsets of $X$.
\\$C(X)$ is the space of real-valued continuous functions on $X$.
\\ $C_b(X)$ is the space of bounded functions $f\in C(X)$.
\\$\LL(X)$ is the space of functions $f\in C_b(X)$ such that
$$
\|f\|_{\LL}:=\|f\|_\infty+\sup_{u\neq
v}\frac{|f(u)-f(v)|}{\|u-v\|}<+\infty.
$$$\PP(X)$ is the set of probability measures on $(X,\BBBBB(X)).$
If $\mu\in\PP(X)$ and  $f\in C_b(X)$, we set
$$
(f,\mu)=\int_Xf(u)\mu(\dd u).
$$
If $\mu_1,\mu_2\in\PP(X)$, we set
$$
\|\mu_1-\mu_2\|_\LL^*=\sup\{|(f,\mu_1)-(f,\mu_2)|: f\in\LL(X),
\|f\|_\LL\le1\},
$$
$$
\|\mu_1-\mu_2\|_{var}=\sup\{|\mu_1(\Gamma)-\mu_2(\Gamma)|:\Gamma\in\BBBBB(X)\}.
$$For any $\Gamma_1,\Gamma_2\in\BBBBB(X)$, with $\pP(\Gamma_2)\neq0$, denote
$$
\pP(\Gamma_1|\Gamma_2)=\frac{\pP(\Gamma_1\Gamma_2)}{\pP(\Gamma_2)}.
$$
The distribution of a random variable $\xi$ is denoted by
$\DD(\xi)$. We denote by $C, C_k$ unessential positive constants.
\section{Preliminaries}\label{aprest}
It is well known that  problem (\ref{E:eq})-(\ref{E:ic}) with $\eta
\equiv 0$ and $u_0\in H$ has a unique solution in the space $C(\R_+,
H)\cap L^2_{\text{loc}}(\R_+,H^2(D))$. Let $S_t:H\rightarrow H$ be
the resolving semi-group for that problem. Let $\tau_0\equiv 0$ and
define $u_t$  by the  relation
\[
u_t=
\begin{cases}
S_{t-\tau_k}(u_{\tau_k}), &\text{if $t\in
[\tau_k,\tau_{k+1})$},\,\,\, k\ge
0,\\S_{t_{k+1}}(u_{\tau_{k}})+\eta_{k+1},      &\text{if
$t=\tau_{k+1}$}.
\end{cases}
\]
Then $u_t$ is the unique solution of  problem
(\ref{E:eq})-(\ref{E:pr}). Clearly, $u_t$ exists for all $t>0$ with
probability 1, as $\pP\{\sum t_k=\infty\}=1$. Let us define a
continuous functional on $H$:
\begin{equation}\label{E:H}
\HH(u)=\int_D\Big(\alpha|\nabla
u(x)|^2+\frac{\beta}{4}|u(x)|^4\Big)\dd x,
\end{equation}where $\alpha$ is a positive constant. If $\alpha$ is sufficiently small,  we have the estimate
\begin{align}
\HH(S_t(u))\le e^{-at}\HH(u),\,\,\,\,\,t\ge0,\label{E:gn1}
\end{align}where $a$ is a positive constant, and there is a constant $C$ such that
\begin{align}
&\|S_t(u)-S_t(v)\|_1\le C\exp(C(\|u\|_1^6+\|v\|_1^6))\|u-v\|_1,\,\,\,t\ge0, \label{E:n1}\\
&\|S_t(u)-S_t(v)\|_2\le
Ct^{-\frac{1}{2}}\exp(C(\|u\|_1^6+\|v\|_1^6))\|u-v\|_1,\,\,\,t>0,\label{E:gn3}
\end{align}
 where  $u,v\in H$. The proof (\ref{E:gn1}), (\ref{E:n1}) and (\ref{E:gn3}) is carried out by standard methods and is given in the Appendix.\\
For any sequence $a_k$, $m\le k\le n,$ we set
$$
\langle a_k\rangle_m^n=\frac{1}{n-m+1}\sum_{k=m}^{n}a_k.
$$
Suppose $u_k, u_k'\in H$ and $t_k>0$ are arbitrary sequences. Define
$\zeta_k$ and $\zeta_k'$ by the  relations
\begin{equation}\label{E:ze}
u_k=S_{t_k}(u_{k-1})+\zeta_k,\,\,\,\,\,\,u_k'=S_{t_k}(u_{k-1}')+\zeta_k'.
\end{equation}
\begin{lemma}\label{L:P}
Suppose that
\begin{align}
&P_N u_k=P_Nu_k',\,\,\,\,\,Q_N\zeta_k=Q_N\zeta_k',\,\,\,\,\,\,\,\,\,l+1\le k\le n,\label{E:Y}\\
&e_j\in H_N,\,\,\,\,\,\,\,\,\,\,j=1,...,N' \label{E:Y1}
\end{align}
for some $N'\ge1$ and  $N\ge1$.
 Then
\begin{align}\label{E:Q1}
\|u_k-u_k'\|_1&\le
(C\alpha_{N'+1}^{-\frac{1}{2}})^{k-l}\Big(\prod_{i=l+1}^kt_i\Big)^{-\frac{1}{2}}\nonumber\\&\times\exp\big(C(k-l)(\langle\|u_i\|_1^6\rangle_l^{k-1}+\langle\|u_i'\|_1^6\rangle_l^{k-1})\big)\|u_l-u_l'\|_1,
\end{align}for $l\le k\le n,$ where $C$ is a positive constant not depending on $u_k, u_k',n,l,N$ and $N'$.
\end{lemma}
\begin{proof} Using (\ref{E:gn3}), (\ref{E:Y}) and (\ref{E:Y1}), we see that
\begin{eqnarray}
\|u_k-u_k'\|_1&=&\|Q_N(u_k-u_k')\|_1=\|Q_N(S_{t_k}(u_{k-1})-S_{t_k}(u_{k-1}'))\|_1\nonumber\\&\le&\|Q'_{N'}(S_{t_k}(u_{k-1})-S_{t_k}(u_{k-1}'))\|_1\nonumber\\&\le&\alpha_{N'+1}^{-\frac{1}{2}}\|S_{t_k}(u_{k-1})-S_{t_k}(u_{k-1}')\|_2\nonumber\\&\le&
C
\alpha_{N'+1}^{-\frac{1}{2}}t_k^{-\frac{1}{2}}\exp\big(C(\|u_{k-1}\|_1^6+\|u_{k-1}'\|_1^6)\big)\|u_{k-1}-u_{k-1}'\|_1\nonumber.
\end{eqnarray}
Iteration of this inequality results in (\ref{E:Q1}).
\end{proof}
\section{Markov chains associated with CGL equation and existence of stationary measures}\label{S:MC}
Let $u_0$ be an $H$-valued random variable, independent of
$\{\eta_k\}$ and $\{t_k\}$, and let $u_t$ be the solution of problem
(\ref{E:eq})-(\ref{E:pr}). Denote by $\FF_t$, $t\ge0$ the
$\sigma$-algebra generated by $u_0$ and $\{\zeta(s), 0\le s\le t
\}$, where
\begin{equation}\label{E:zeta}
\zeta(s)=\sum_{k=1}^\infty I_{\{\tau_k\le s\}}\eta_k.
\end{equation}
\begin{lemma}\label{L:MP} Under the above conditions, $u_t$ is a homogeneous Markov process with respect to  $\FF_t$.
\end{lemma}
The proof of this lemma is given in the Appendix. For any $u\in H$
and $\Gamma\in \BBBBB (H)$, we set
$P_t(u,\Gamma)=\pP\{u_t(u)\in\Gamma\}.$ The Markov operators
corresponding to the process $u_t$ have the form
$$
\PPPP_t f(u)=\int_HP_t(u,\dd
v)f(v),\,\,\,\,\,\,\,\PPPP^*_t\mu(\Gamma)=\int_H
P_t(u,\Gamma)\mu(\dd u),
$$ where $f\in C_b(H)$ and $\mu\in\PP(H).$

The strong Markov property implies that $u_{\tau_k}$ is a
homogeneous Markov chain with respect to $\sigma$-algebra $\GG_k$
generated by $\{\eta_n, t_n,1\le n\le k\}$. In what follows, we
shall write $u_k$ instead of $u_{\tau_k}$; this will not lead to
confusion.
\begin{lemma}
\begin{enumerate}
\item[(i)] For any $\e>0$ there is a constant $C_\e>0$ such that
\begin{equation}\label{E:gnk}
\HH(u_k)\le (1+\e)^k e^{-a\tau_k}\HH(u_0)+C_\e\sum_{l=1}^k
e^{-a(\tau_k-\tau_l)}(1+\e)^{k-l}\HH(\eta_l).
\end {equation}
\item[(ii)] Let $\E\HH(\eta_k)^p<\infty$ for some  $p\ge 1$. Then
\begin{equation}\label{E:gnk2}
\E\HH(u_k)^p\le\gamma^k\E\HH(u_0)^p+\frac{C_p}{1-\gamma}\E\HH(\eta_k)^p,
\end{equation} where $0<\gamma<1$ and $C_p>0$ are some constants not depending on $k.$
\end{enumerate}
\end{lemma}
\begin{proof}
Using (\ref{E:gn1}), we obtain
$$
\HH(u_k)\le(1+\e)e^{-t_k a}\HH(u_{k-1})+C_\e\HH(\eta_k).
$$Iteration of this inequality results in (\ref{E:gnk}).\\ To prove (\ref{E:gnk2}), note that for any $\e>0$ there is a constant $C_{p,\e}$ such that
\begin{equation}\label{E:gnk3}
\HH(u_k)^p\le(1+\e)e^{-t_k ap}\HH(u_{k-1})^p+C_{p,\e}\HH(\eta_k)^p.
\end{equation}Taking the expectation and using the independence of $t_k$ and $u_{k-1}$, we obtain
$$
\E\HH(u_k)^p\le(1+\e)\frac{\lambda}{\lambda+ap}\E\HH(u_{k-1})^p+C_{p,\e}\E\HH(\eta_k)^p.
$$ Choosing $\e>0$ so small that $\gamma:=(1+\e)\frac{\lambda}{\lambda+ap}<1$ and iterating the resulting inequality, we arrive at (\ref{E:gnk2}).
\end{proof}
\begin{lemma}\label{L:T}
Let $\E\|\eta_k\|_1^p<\infty$ for all $p\ge1$ and let $u_0\in H$.
Then
\begin{enumerate}
\item[(i)] There is a constant $M>0$ not depending on $u_0$ and a random integer $T=T(u_0)\ge1$ such that
\begin{align}
&\langle\|u_k\|_1^6\rangle_0^n\le M \,\,\,\,\,\,\,\,\text{for  $n\ge T$,}\label{E:T1}\\
&\E T^p<\infty\,\,\,\,\,\,\,\,\,\,\,\,\,\,\,\,\,\,\,\,\text{for all
$p\ge1$.}\label{E:T2}
\end{align}
\item[(ii)] For any $\delta\in(0,1)$ and $d>0$, there is a constant $R=R(\delta, d)>0$ such that
\begin{equation}
\pP\{\langle\|u_k\|_1^6\rangle_0^n\le R,\forall
n\ge0\}\ge\delta,\label{E:T3}
\end{equation}for any  $u_0\in B_d$, where $B_d=\{u\in H:\|u\|_1\le d\}$.
\end{enumerate}
\end{lemma}
\begin{proof}
Let us fix $\e>0.$ Using (\ref{E:gnk3}) with $p=3$, we obtain
\begin{align}\label{E:O1}
&\HH(u_k)^3\le (1+\e) e^{-3at_k}\HH(u_{k-1})^3+C_\e
\HH(\eta_k)^3\nonumber\\&=(1+\e)\Big(
e^{-3at_k}-\frac{\lambda}{\lambda+3a}\Big)\HH(u_{k-1})^3+
(1+\e)\frac{\lambda}{\lambda+3a}\HH(u_{k-1})^3+C_\e \HH(\eta_k)^3.
\end{align} Choosing $\e>0$ so small that $q:=(1+\e)\frac{\lambda}{\lambda+3a}<1$ and summing up inequalities (\ref{E:O1}) for $1\le k\le n,$ we arrive at
\begin{align}
\sum_{k=1}^n\HH(u_k)^3&\le(1+\e)\sum_{k=1}^n\Big(
e^{-3at_k}-\frac{\lambda}{\lambda+3a}\Big)\HH(u_{k-1})^3+q\sum_{k=1}^n\HH(u_{k-1})^3\nonumber\\&+C_\e\sum_{k=1}^n\HH(\eta_k)^3,\nonumber
\end{align} whence it follows that
\begin{align}\label{E:kargn}
\alpha^3\langle\|u_k\|_1^6\rangle_0^n&\le\langle\HH(u_k)^3\rangle_0^n\le\frac{1+\e}{1-q}\frac{1}{n+1}\sum_{k=1}^n\Big(
e^{-3at_k}-\frac{\lambda}{\lambda+3a}\Big)\HH(u_{k-1})^3\nonumber\\&+\frac{1}{1-q}\frac{\HH(u_0)^3}{n+1}+\frac{C_\e}{1-q}\mmm+\frac{C_\e}{1-q}\frac{1}{n+1}\sum_{k=1}^n(\HH(\eta_k)^3-\mmm),
\end{align} where $\mmm=\E\HH(\eta_k)^3.$
To complete the proof, we need the following lemma, whose proof is
given in the Appendix.
\begin{lemma}\label{L:Ap}
Suppose that $M_k$ is a sequence of random variables that satisfies
the inequality
\begin{equation}\label{E:Mk}
\E|M_k|^{2p}\le C_pk^p\,\,\,\,\,\,\,\,\,\,\,\,\text{for all
$p\ge1.$}
\end{equation}  Then the following assertions take place.
\begin{enumerate}
\item[(i)] There is a random integer $T\ge1$ such that
\begin{align}
&\frac{1}{k}|M_k|\le 1 \,\,\,\,\,\,\,\text{for $k\ge T,$}\label{E:Ap1}\\
&\E T^p<\infty \,\,\,\,\,\,\,\,\,\text{for all
$p\ge1.$}\label{E:Ap2}
\end{align}
\item[(ii)] For any $\delta\in(0,1)$, there is a constant $R>0$ such that
\begin{equation}\label{E:Mkgn}
\pP\{\frac{|M_k|}{k}\le R,\forall k\ge1\}\ge\delta.
\end{equation}
\end{enumerate}
\end{lemma}  Let us set
\begin{align}
&M_k'=\sum_{i=1}^k\Big( e^{-3at_i}-\frac{\lambda}{\lambda+3a}\Big)\HH(u_{i-1})^3,\nonumber\\
&M_k''=\sum_{i=1}^k(\HH(\eta_i)^3-\mmm),\,\,\,\,\,\,M_0'=M_0''=0.\nonumber
\end{align} Clearly $M_k'$ and $M_k''$ are martingales. For $M_k''$ it is easy to verify that (\ref{E:Mk}) holds, as $M_i''-M_{i-1}''=\HH(\eta_i)^3-\mmm$, $i\ge1$ are centered  i.i.d. random variables. To prove (\ref{E:Mk}) for $M_k'$, we need  Burkholder's inequality for martingales (\cite{H}, Section 2.4):
\begin{equation}\label{E:B}
C_1 \E\Big|\sum_{i=1}^k X_i^2\Big|^p\le\E|M_k'|^{2p}\le C_2
\E\Big|\sum_{i=1}^k X_i^2\Big|^p,
\end{equation}where $X_i=M_i'-M_{i-1}'$, $i\ge1$, $p\ge1$ and $C_1, C_2$ are positive constants depending only on $p$. Using (\ref{E:B}) and (\ref{E:gnk2}), we obtain
\begin{eqnarray}
\E|M_k'|^{2p}&\le& C_2\E \Big|\sum_{i=1}^k\Big(
e^{-3at_i}-\frac{\lambda}{\lambda+3a}\Big)^2\HH(u_{i-1})^6\Big|^p\nonumber\\&\le&
C \sum_{i=1}^k \E\HH(u_{i-1})^{6p}{k^{p-1}}\le C'k^p,\nonumber
\end{eqnarray}where $C'$ depends on $\|u_0\|_1$. Applying Lemma \ref{L:Ap}, let $T'$ and $T''$ be the random variables corresponding to martingales $M_k'$ and $M_k''$. Setting
$$
T=T_1\vee
T_2\vee\Big(\HH(u_0)\frac{1}{1-q}\Big),\,\,\,\,M=\Big(\frac{1+\e}{1-q}+\frac{C_\e(\mmm+1)}{1-q}+1\Big)\alpha^{-3},
$$ it is easy to verify that we have (\ref{E:T1}) and (\ref{E:T2}) for $T$ and $M$.
\\ To prove (\ref{E:T3}), we apply (\ref{E:Mkgn}) to the sequence
$$
M_k=\frac{1+\e}{1-q}M_k'+\frac{C_\e}{1-q}M_k'',
$$ and using (\ref{E:kargn}), we  see that (\ref{E:T3}) holds with
$$
R_1=R+C_d\frac{1}{1-q}+\frac{C_\e}{1-q}\mmm,
$$
where $C_d=\sup_{u\in B_d}\HH(u)^3.$\end{proof} Let $\tau_R$ be the
first hitting time of the ball $B_R$:
$$
\tau_R=\min\{k\ge0:\|u_k\|_1\le R\}.
$$
\begin{lemma}\label{L:(i)}
Let $\E\HH(\eta_1)<+\infty$. Then there are positive constants
$\delta$, $C$ and $R$ not depending on $u$ such that
$$
\E_u e^{\delta\tau_R}\le C (1+\HH(u)).
$$
\end{lemma}\begin{proof}
It suffices to show that $u_k$ possesses a Lyapunov function (see
\cite{MT}), i.e. there is a continuous functional $F$ on $H$ such
that
\begin{enumerate}
\item[(i)]$F(u)\ge1$ and $\lim_{\|u\|_1\rightarrow\infty}F(u)=+\infty.$
\item[(ii)] There are positive constants $n,R',C'$ and $a<1$ that
\begin{align}
&\E_uF(u_n)\le aF(u)\,\,\,\,\,\,\text{for $\|u\|_1\ge R'$,}\label{E:L11}\\
&\E_uF(u_k)\le C'\,\,\,\,\,\,\,\,\,\,\,\,\,\,\,\,\text{for $\|u\|_1<
R'$, $k\ge0.$}\label{E:L22}
\end{align}
\end{enumerate}
Let
\[
F(u)=
\begin{cases}
\HH(u), &\text{if $\HH(u)\ge A$} ,\\A,      &\text{if $\HH(u)<A$},
\end{cases}
\] where $A\ge 1$. Then (i) is satisfied. Let $\|u\|_1\ge R'.$ Note that
\begin{align}
\E_uF(u_n)&=\E_uF(u_n)I_{\{\HH(u_n)<A\}}+\E_uF(u_n)I_{\{\HH(u_n)\ge
A\}}\nonumber\\&\le A+\E_u\HH(u_n)\le
\gamma^n\HH(u)+A+C\E\HH(\eta_1),
\end{align}where we used (\ref{E:gnk2}). Choosing $n$ and $R'$ so large that $2\gamma^n<1$ and $A+C\E\HH(\eta_1)\le\gamma^nR'^2\alpha$, where $\alpha$ is the constant in (\ref{E:H}), we arrive at (\ref{E:L11}) with $a=2\gamma^n.$ It remains to note that (\ref{E:L22}) follows from (\ref{E:gnk2}).
 \end{proof}
\begin{definition}
A measure $\mu\in\PP(H)$ is said to be stationary for problem
(\ref{E:eq}), (\ref{E:lc}), (\ref{E:pr}), if $\PPPP_t^*\mu=\mu$ for
any $t\ge0$.
\end{definition}Using the  classical Bogolyubov-Krylov argument and Fatou's lemma,  one can prove the following
theorem. Its proof is outlined in the Appendix.
\begin{theorem}\label{T:BK}
Let $\E\HH(\eta_k)<\infty,$ then  problem (\ref{E:eq}),
(\ref{E:lc}), (\ref{E:pr}) has at least one stationary measure.
Moreover, if $\E\HH(\eta_k)^p<\infty$ for some $p\ge1,$ then  for
any stationary measure $\mu$ we have:
$$
\HH_p(\mu):=\int_H\HH(u)^p\mu(\dd u)<+\infty.
$$
\end{theorem} We denote by $\PP_1(H)$ the set of measures $\mu\in\PP(H)$ such that $\HH(\mu):=\HH_1(\mu)<+\infty.$

\section{Main result}\label{MR}
To show the uniqueness of  stationary measure for  (\ref{E:eq}),
(\ref{E:lc}), (\ref{E:pr}), we shall need the following condition
satisfied for $\eta_k$:
\begin{condition}\label{C:1}
The random variables $\eta_k$ are i.i.d. and have the form
$$
\eta_k=\sum_{j=1}^\infty b_j\xi_{jk}g_j(x),
$$ where $\{g_i\}_{i\in\N}$ is an orthonormal basis in $H$, $b_j\ge0$ are some constants with
$$
B:=\sum^\infty_{j=1} b_j^2<\infty,
$$ and  $\xi_{jk}$ are independent scalar random variables. Moreover, the distribution of $\xi_{jk}$ possesses a density $p_j(r)$ (with respect to the Lebesgue measure), which is a function of bounded variation such that
\begin{equation}\label{E:paj}
 \int_{-\e}^{\e}p_j(r)\dd r>0,\,\,\,\,\int_{-\infty}^{+\infty}|r|^p p_j(r)\dd r\le C_p<\infty,
\end{equation} for all $\e>0$, $p\ge1$, $j\ge1$ and for some constants $C_p>0$.
\end{condition}
Clearly, if Condition \ref{C:1} is satisfied, then
\begin{equation}\label{E:sah}
\E\|\eta_k\|_1^p<\infty\,\,\,\,\text{for all $k\ge1$, $p\ge1$}.
\end{equation}
\begin{theorem}\label{T:him}
Suppose that Condition \ref{C:1} is satisfied. For any $B>0$ there
is an integer $N'\ge1$ such that, if
\begin{equation}\label{E:HN*}
e_j\in H_N,\,\,\,\,\,\,j=1,...,N'
\end{equation}for some $N\ge1$, and
\begin{equation}\label{E:HN**}
b_j\neq0,\,\,\,\,\,\,\,\,\,\, j=1,...,N,
\end{equation}
then there is a unique stationary measure $\mu\in\PP(H).$ Moreover,
for any initial measure $\mu'\in\PP_1(H)$ we have
\begin{equation}\label{E:him}
\|\PPPP_t^*\mu'-\mu\|_\LL^*\le C_p(1+\HH(\mu'))t^{-p},\,\,\,\, t>0,
\end{equation}   where $C_p$ is a constant not depending on $\mu'$.
\end{theorem}
\begin{proof} \textbf{ Step 1.} It suffices to show that for any $u,u'\in H$ we have
\begin{equation}\label{E:h22}
|\PPPP_tf(u)-\PPPP_tf(u')|\le C_p\|f\|_\LL(1+\HH(u)+\HH(u'))t^{-p},
\end{equation}for any $p\ge1,$ $t>0$ and some constant $C_p>0$ not depending on $(u,u')$ and $t$.
\\Indeed, suppose that (\ref{E:h22}) is already proved. Then for any two initial measures $\mu',\mu''\in \PP_1(H)$ we derive from  (\ref{E:h22}):
 \begin{equation}\label{E:h11}
\|\PPPP_t^*\mu'-\PPPP_t^*\mu''\|_\LL^*\le
C_p(1+\HH(\mu')+\HH(\mu''))t^{-p}.
\end{equation}This inequality shows the uniqueness of stationary measure in $\PP_1(H)$. It follows from Theorem \ref{T:BK} that any  stationary measure $\mu$ is in $\PP_1(H)$. Taking $\mu''=\mu$ in (\ref{E:h11}), we arrive at (\ref{E:him}).
\\\textbf{ Step 2.} Inequality (\ref{E:h22}) is a direct consequence of the following proposition.
\begin{proposition}\label{P:A}
Under the conditions of Theorem \ref{T:him}, for any $B>0$ there is
an integer $N'\ge1$ such that, if (\ref{E:HN*}) and  (\ref{E:HN**})
hold for some integer $N\ge1$, then there is a probability space
$(\Omega,\FF,\pP)$ and a sequence of i.i.d. random variables
$\{t_k\}$ that are exponentially distributed with parameter $\la$
such that for any $u,u'\in H$ one can construct random sequences
$u_k,u_k'$ defined on $\Omega$ with the following properties:
\begin{enumerate}
\item[(i)] The initial value of the trajectory $(u_k,u_k')$ is $(u,u')$:
 $$
u_0=u,\,\,\,u_0'=u'.
$$
Furthermore, the random variables $\zeta_k$ and $\zeta_k'$ defined
by (\ref{E:ze}) are i.i.d., and their distribution coincides with
that of $\eta_k$:
$$
\DD(\zeta_k)=\DD(\zeta_k')=\DD(\eta_k).
$$
\item[(ii)] There is a random integer $\ell=\ell(u,u')$ and a constant $M$ depending only on $B$ such that
\begin{align}
&P_N u_k=P_N u_k' &\text{for $k\ge\ell+1,$}\label{E:P1}\\
&Q_N\zeta_k=Q_N\zeta_k' &\text{for $k\ge1,\,\,\,\,\,\,\,\,\,\,$}\label{E:P2}\\
&\langle\|u_i\|_1^6+\|u_i'\|_1^6\rangle_\ell^k\le M &\text{for $k\ge\ell+1,$}\label{E:P3}\\
&\frac{1}{2}\frac{1}{k-\ell}\big|\sum_{i=\ell+1}^{k}\log t_i\big|\le
M&\text{for $k\ge\ell+1$}\label{E:P4}.
\end{align}
\item[(iii)] There is a positive constant $C_p$ not depending on $(u,u')$ such that
\begin{align}
&\E \ell^p\le C_p (1+\HH(u)+\HH(u'))&\text{for all $p\ge1,$}\label{E:P5}\\
&\|u_\ell\|_1\vee \|u_\ell'\|_1\le 1.\label{E:P6}
\end{align}
\end{enumerate}
\end{proposition}
To prove (\ref{E:h22}), let $u_k$ and $u_k'$ be the random sequences
constructed in Proposition \ref{P:A} and  corresponding to the
initial value $(u,u').$ Let $\tau_k=\sum_{n=1}^kt_n,n\ge1$ and
$\tau_0=0.$ Define
\[
u_t=
\begin{cases}
S_{t-\tau_k}(u_k), &\text{if $t\in [\tau_k,\tau_{k+1})$},\,\,\, k\ge
0,\\S_{t_{k+1}}(u_k)+\zeta_{k+1},      &\text{if $t=\tau_{k+1}$},
\end{cases}
\]and $u'_t$ is defined in a similar way. Clearly, $u_t$ and $u_t'$ have the same distributions as the solutions of (\ref{E:eq})-(\ref{E:pr}) corresponding to $u$ and $u'$, respectively. Thus
\begin{equation}\label{E:bet}
|\PPPP_t f(u)-\PPPP_t f(u')|=|\E(f(u_t)-f(u_t'))|.
\end{equation}Let
\begin{equation}\label{E:Nt}
\NN_t=\max\{k\ge0:\tau_k\le t\},
\end{equation} then $\NN_t$ is a Poisson random variable with parameter $\la t$ (e.g., see \cite{Kar}). Define $G_t=\{\omega: 2\ell+1\le \NN_t\}=\{\omega:\tau_{2\ell+1}\le t\}$. As $\tau_k$ is a Gamma random variable with parameters $\la$ and $k$ (e.g., see \cite{Fel}), we have\begin{align}
\E\tau_{2\ell+1}^q&\le\sum_{n=1}^\infty\E[\tau_{2n+1}^qI_{\{\ell=n\}}]\le
\sum_{n=1}^\infty(E\tau_{2n+1}^{2q})^\frac{1}{2}\pP\{\ell=n\}^\frac{1}{2}\nonumber\\&\le
C(1+\HH(u)+\HH(u'))\sum_{n=1}^\infty\frac{1}{n^2}<\infty,
\end{align}for any $q\ge1$, where we used the Cauchy--Schwarz inequality and (\ref{E:P5}) with $p=2(2+q)$. It follows that
\begin{equation}\label{E:P7}
\pP(G_t^c)\le C_p'(1+\HH(u)+\HH(u'))t^{-p}\,\,\,\,\,\,\,\text{for
any $p\ge1$}.
\end{equation} Using (\ref{E:n1}), we see that
$$
\|u_t-u_t'\|_1\le
C\exp(C(\|u_{\tau_{\NN_t}}\|_1^6+\|u'_{\tau_{\NN_t}}\|_1^6))\|u_{\tau_{\NN_t}}-u'_{\tau_{\NN_t}}\|_1,
$$ whence, using (\ref{E:P1})-(\ref{E:P4}), (\ref{E:P6}) and Lemma \ref{L:P}, we obtain
\begin{equation}
\E[I_{G_t}\|u_t-u_t'\|_1]\le\E[2(C\alpha_{N'+1}^{-\frac{1}{2}})^{\NN_t-\ell}e^{2CM(\NN_t-\ell)}]\nonumber.
\end{equation}Choosing $N'$ so large that $\log \alpha_{N'+1}\ge2(2CM+\log C+2)$, we arrive at
\begin{equation}\label{E:P8}
\E[I_{G_t}\|u_t-u_t'\|_1]\le\E e^{-\NN_t}=e^{-ct},
\end{equation}where $c=\la-\frac{\la}{e}$.
Let $f\in\LL(H).$ Then, using (\ref{E:bet}), (\ref{E:P7}) and
(\ref{E:P8}), we derive
\begin{eqnarray}
|\PPPP_tf(u)-\PPPP_tf(u')|&\le&\E|f(u_t)-f(u_t')|\nonumber\\&\le&\|f\|_\LL\E[I_{G_t}\|u_t-u_t'\|_1]+\E
I_{G_t^c}2\|f\|_{\infty}\nonumber\\&\le&\|f\|_\LL e^{-ct}+2\|f\|_\LL
C_p'(1+\HH(u)+\HH(u'))t^{-p}\nonumber\\&\le&
C_p\|f\|_\LL(1+\HH(u)+\HH(u'))t^{-p}.
\end{eqnarray}This completes the proof of (\ref{E:h22}).
\end{proof}
\begin{remark}
The embedded Markov chain $u_{\tau_k}$ also satisfies a property of
polynomial mixing. This follows from Proposition \ref{P:A} and is
proved using the same arguments as in the proof of Theorem
\ref{T:him}. The stationary measures of the original process and
that of embedded chain are connected with the Khasminskii relation:
$$
(f,\mu)=\frac{1}{\E_\nu\tau_1}\E_\nu\int_0^{\tau_1}f(u_t)\dd t,
$$  where $\nu$ and $\mu$ are the stationary measures of $u_{\tau_k}$ and $u_t$ respectively.
\end{remark}

\section{Coupling operators}\label{CO}

Let $\eta_k$ be a sequence of random variables with range in $H$ and
suppose that Condition \ref{C:1} is satisfied for $\eta_k.$ Clearly,
if $b_j\neq0$,  $j=1,...,N,$ then the distribution of the random
variable $P_N(\eta_1)$ is absolutely continuous with respect to the
Lebesgue measure, and its density has the form
$$
p(x):=\prod_{j=1}^N
q_j(x_j),\,\,\,\,q_j(x_j)=b_j^{-1}p_j(x_jb_j^{-1}),\,\,\,x=(x_1,...,x_N)\in
H_N.
$$ Now we have the following lemma, which is a version of Lemma 3.2 in \cite{KSH}:
\begin{lemma}\label{L:cp}
Suppose that  Condition \ref{C:1} is satisfied and $b_j\neq0$ for
$j=1,...,N,$ where $N\ge 1$ is an integer. Then there is a
probability space $(\Omega, \FF,\pP)$ such that for any $u,u'\in H$
there are $H$-valued random variables $\zeta=\zeta(u,u',\omega)$,
$\zeta'=\zeta'(u,u',\omega)$ and a  real-valued random variable
$t=t(\omega)$ with the following properties:
\begin{enumerate}
\item[(i)] The random variables $\zeta,\zeta'$ and $\eta_1$ have the same distributions, and $t$ is exponentially distributed with parameter~$\lambda.$
\item[(ii)] The random variables $(P_N\zeta,P_N\zeta')$ and $(Q_N\zeta, Q_N\zeta')$ are independent, and $\zeta$ and $\zeta'$ are independent of $t$.
\item[(iii)] The random variables $Q_N\zeta$ and $Q_N\zeta'$ are equal for all $\omega\in\Omega$ and do not depend on $(u,u').$
\item[(iv)]  The random variables $\zeta$ and $\zeta'$ are measurable functions of  $(u,u',\omega)\in H\times H\times \Omega.$
\end{enumerate}
\end{lemma}
\begin{proof}
Suppose that $t_1=t_1(\omega_1)$ is a random variable that is
exponentially distributed with parameter $\lambda$ and is  defined
on the  space $(\Omega_1,\FF_1,\pP_1)$. Let $(v,v')$ be a maximal
coupling for $(\nu_{u,\omega_1},\nu_{u',\omega_1})$, where
$\nu_{u,\omega_1}$ is a measure on $H_N$ given by the density
$p(x-P_NS_{t_1(\omega_1)}(u))$ (see \cite{Li}, Section I, 5). By
Theorem 4.2 in \cite{KSH}, we can assume that the random variables
$v$ and $v'$ are defined on the same probability space
$(\Omega_2,\FF_2,\pP_2)$ for all $u,u'\in H$, $\omega_1\in\Omega_1$
and are measurable functions of $(u,u',\omega_1,\omega_2)\in H\times
H\times\Omega_1\times\Omega_2$. Suppose that $\eta_1$ is defined on
the space $(\Omega_3,\FF_3,\pP_3)$. We denote by $(\Omega, \FF,\pP)$
the direct product of $(\Omega_i,\FF_i,\pP_i)$, $i=1,2,3,$ and
define $\zeta$, $\zeta'$ and $t$ by the relations:
\begin{align}
t(\omega)&=t_1(\omega_1)\nonumber,\\
P_N\zeta(\omega)&=v(u,u',\omega_1,\omega_2)-P_NS_{t(\omega_1)}(u)\nonumber,\\P_N\zeta'(\omega)&=v'(u,u',\omega_1,\omega_2)-P_NS_{t(\omega_1)}(u')\nonumber,\\
Q_N\zeta(\omega)&=Q_N\zeta'(\omega)=Q_N\eta_1(\omega_3),\nonumber
\end{align}
where $\omega=(\omega_1,\omega_2,\omega_3)\in\Omega $. Using the
definition of $\zeta$ and Fubini's theorem, we see that
\begin{align}\label{E:label}
\pP\{P_N\zeta\in\Gamma\}&=\E
I_{\{P_N\zeta\in\Gamma\}}=\E_1\pP_2\{P_N\zeta(\omega_1)\in\Gamma\}\nonumber\\&=\E_1\pP_2\{v-P_NS_{t_1}(u)\in
\Gamma\}=\int_\Gamma p(x)\dd x =\pP\{P_N\eta_1\in\Gamma\},
\end{align} for any $\Gamma\in\BB(H_N)$, where $\E_1$ is the expectation corresponding to the measure~$\pP_1$. All assertions of lemma follow from the construction and relation (\ref{E:label}).
\end{proof}
\begin{remark}\label{R:2}
Using inequality (3.8) in  Lemma 3.2, \cite{KSH} for the variational
distance between $\nu_{u,\omega_1}$ and $\nu_{u',\omega_1}$, we
obtain the inequality:
$$
\|\nu_{u,\omega_1}-\nu_{u',\omega_1}\|_{var}\le
C_N\|S_t(u)-S_t(u')\|_1,
$$ which holds $\pP_1$-a.s.. Then the  definition of maximal coupling gives
\begin{equation}\label{E:kar}
\pP_2\{v\neq v'\}\le C_N\|S_t(u)-S_t(u')\|_1.
\end{equation}
\end{remark}
\begin{remark}\label{R:1}
Let $(\Omega',\FF',\pP')$ be the direct product of
$(\Omega_i,\FF_i,\pP_i),$ $i=2,3.$ For any $\omega_1\in\Omega_1$,
let $E_{\omega_1}=\{\omega'\in\Omega':v(u,u',\omega_1,\omega')\neq
v'(u,u',\omega_1,\omega')\}$. As $(v,v')$ is a maximal coupling for
$(\nu_{u,\omega_1},\nu_{u',\omega_1})$, we have
\begin{align}
&\pP'\{v(u,u',\omega_1,\cdot)\in\Gamma,v'(u,u',\omega_1,\cdot)\in\Gamma'|E_{\omega_1}\}\nonumber\\\nonumber&=\pP'\{v(u,u',\omega_1,\cdot)\in\Gamma|E_{\omega_1}\}\pP'\{v'(u,u',\omega_1,\cdot)\in\Gamma'|E_{\omega_1}\},
\end{align} if $\pP'\{E_{\omega_1}\}>0$ and $\Gamma,\Gamma'\in\BB(H).$ Now it is easy to notice that
\begin{align}\label{E:pp}
\pP'\{v_{\omega_1}\in\Gamma,v'_{\omega_1}\in\Gamma',E_{\omega_1}\}\ge\pP'\{{v_{\omega_1}\in\Gamma,E_{\omega_1}\}}\pP'\{v'_{\omega_1}\in\Gamma',E_{\omega_1}\}.
\end{align}
\end{remark}
Let us define coupling operators by the formulas
$$
\RR(u,u',\omega)=S_{t(\omega)}(u)+\zeta(u,u',\omega),\,\,\,\RR'(u,u',\omega)=S_{t(\omega)}(u')+\zeta'(u,u',\omega),
$$ where  $u, u'\in H$ and $\omega\in\Omega$.
\begin{lemma}\label{L:2}
Under the conditions of Lemma \ref{L:cp}, there exists a constant
$\gamma\in(0,1)$ such that  for any $r>0$ and an appropriate
constant $\e:=\e(r)>0$ we have
\begin{equation}\label{E:h1}
P^1_r:=\pP\{\HH(\RR(u,u',\cdot))+\HH(\RR'(u,u',\cdot))\le\big(\gamma(\HH(u)+\HH(u'))\big)\vee
r\}>\e,
\end{equation}
for all  $u,u'\in H.$
\end{lemma}
\begin{proof} \textbf{ Step 1.} It suffices to show that there is $C>0$ such that for any $\delta>0$ and an appropriate constant $\e_\delta>0$ the following inequality holds $\pP_1$-a.s.:
\begin{align}\label{E:h2}
P_\delta^2(\omega_1):=\pP'\{&\HH(\RR(u,u',\omega_1,\cdot))+\HH(\RR'(u,u',\omega_1,\cdot))\nonumber\\&\le
C(\HH(S_{t(\omega_1)}(u))+\HH(S_{t(\omega_1)}(u')))+\delta\}\ge
\e_\delta.
\end{align}
Indeed, define the event
$$
V=\{e^{-at(\omega_1)}\le (2C)^{-1}\}.
$$Then $\pP_1(V)>0$, as $t$ is exponentially distributed. We deduce from (\ref{E:gn1}):
\begin{align}
 C(\HH(S_{t(\omega_1)}(u))+\HH(S_{t(\omega_1)}(u')))+\delta&\le  Ce^{-at(\omega_1)}(\HH(u)+\HH(u'))+\delta\nonumber\\&\le\frac{1}{2}(\HH(u)+\HH(u'))+\delta\nonumber,
\end{align}if $\omega_1\in V$. Setting $\gamma=\frac{3}{4}$ and $\delta=\frac{r}{3}$, we see that
$$
\frac{1}{2}(\HH(u)+\HH(u'))+\delta\le\big(\gamma(\HH(u)+\HH(u'))\big)\vee
r.
$$Combining this with (\ref{E:h2}), we obtain  $P^1_r\ge \E_1P_\delta^2(\omega_1)I_V(\omega_1)\ge\e_\delta\pP(V)$.
\\\textbf{Step 2.} Let us fix  arbitrary $\delta>0$ and
$\omega_1\in\Omega_1$. Suppose that
\begin{equation}\label{E:ara}
\HH(P_NS_{t(\omega_1)}(u))\le \HH(P_NS_{t(\omega_1)}(u'))
\end{equation} (the proof of the other case is similar). Define the
events
\begin{align}
  &A_\delta=\{\omega'\in\Omega':\HH(P_N\RR(\omega_1,\omega'))\le 8\HH(P_NS_{t(\omega_1)}(u'))+\frac{\delta}{32}\},\nonumber\\
  &F_\delta=\{\omega'\in\Omega':\HH(Q_N\RR(\omega_1,\omega'))\le
  8\HH(Q_NS_{t(\omega_1)}(u))+\frac{\delta}{32}\},\nonumber\\
  &G'_\delta=\{\omega'\in\Omega':\HH(P_N\RR'(\omega_1,\omega'))\le 8\HH(P_NS_{t(\omega_1)}(u'))+\frac{\delta}{32}\},\nonumber\\
  &F'_\delta=\{\omega'\in\Omega':\HH(Q_N\RR'(\omega_1,\omega'))\le
  8\HH(Q_NS_{t(\omega_1)}(u'))+\frac{\delta}{32}\},\nonumber
\end{align} Clearly, if $\omega'\in A_\delta F_\delta$, then
\begin{align}
&\HH(\RR)\le
8\HH(P_N\RR)+8\HH(Q_N\RR)\le64(\HH(P_NS_t(u'))+\HH(Q_NS_t(u)))+\frac{\delta}{2}.\nonumber\end{align}
As $\dim H_N<\infty$, we obtain
$$ \HH(P_NS_t(u))\le C_1 \HH(S_t(u)),
$$ therefore
$$
\HH(Q_NS_t(u))\le C_2 \HH(S_t(u)).
$$Finally, we have
\begin{align}
\HH(\RR(u,u',\omega_1,\omega'))+\HH(\RR'(u,u',\omega_1,\omega'))\le
C(\HH(S_t(u))+\HH(S_t(u')))+\delta,\nonumber
\end{align} if $\omega'\in A_\delta G_\delta'F_\delta F_\delta'.$
Using property (ii) of Lemma \ref{L:cp}, we see that
$$
P^2_\delta(\omega_1)\ge \pP'(A_\delta F_\delta G'_\delta F'_\delta
)=\pP'(A_\delta G'_\delta )\pP'(F_\delta F'_\delta).
$$Hence, it suffices to find a constant $k_\delta>0$ not depending on $\omega_1\in\Omega_1$ such that
\begin{equation}\label{E:kde}
\pP'(A_\delta G'_\delta )\ge k_\delta,\,\,\,\pP'(F_\delta
F'_\delta)\ge k_\delta.
\end{equation}
\textbf{Step 3.} It follows from (\ref{E:paj}) that for any $\tau>0$
there is $q_\tau>0$ such that
\begin{equation}\label{E:o1}
\pP'\{\|\zeta\|_1\le\tau\}\ge
q_\tau,\,\,\,\pP\{\|\zeta'\|_1\le\tau\}\ge q_\tau.
\end{equation} In view of property (iii) of  Lemma \ref{L:cp}, we have
$$
\pP'\{8\HH(Q_N\zeta)=8\HH(Q_N\zeta')\le\frac{\delta}{32}\}\ge
q'_\delta,
$$where $q'_\delta>0$, therefore
$$
\pP'(F_\delta F'_\delta)\ge q'_\delta.
$$
\textbf{Step 4.} We deduce from (\ref{E:o1}) and (\ref{E:ara}) that
\begin{align}\label{E:o2}
\pP'(A_\delta) \ge q'_\delta,\,\,\,\,\,\pP'(G'_\delta) \ge
q'_\delta.
\end{align} Let $E=\{P_N\RR\neq P_N\RR'\}$. Then $A_\delta E^c=G'_\delta E^c=A_\delta G'_\delta E^c$.
If $\pP'(E)=0$, then
$$
\pP'(A_\delta G'_\delta)=\pP'(A_\delta)\ge q_\delta'.
$$ Suppose that $\pP'(E)>0$. Using Remark \ref{R:1}, we obtain
\begin{align}
&\pP'(A_\de G'_\de )=\pP'(A_\de G_\de'E^c)+\pP'(A_\de
G_\de'E)\ge\pP'(A_\de E^c)+\pP'(A_\delta E)\pP'(G'_\delta E).
\end{align}
If $\pP'(A_\delta E^c)\ge\big(\frac{q_\delta'}{2}\big)^2=:
k_\delta$, then $\pP'(A_\delta G_\delta')\ge k_\delta$. If
$\pP'(A_\delta E^c)< k_\delta$, then
$$
\pP'(A_\de G'_\de )\ge(\pP'(A_\delta)-\pP'(A_\delta
E^c))(\pP'(G'_\delta)-\pP'(A_\delta E^c))\ge\Big(q'_\delta
-\Big(\frac{q'_\delta}{2}\Big)^2\Big)^2\ge k_\delta.
$$
This completes the proof of the lemma.
\end{proof}
\section{Proof of Proposition \ref{P:A}}\label{PP}
Let $(\Omega^k,\FF^k,\pP^k)$, $k\ge1$ be independent copies of the
probability space constructed in Lemma \ref{L:cp}, and let
$(\Omega,\FF,\pP)$ be their direct product. Let $u_0=u$ and
$u_0'=u'$, where $u,u'\in H.$ We set
\begin{align}
&u_k(\omega)=\RR(u_{k-1}(\omega),u'_{k-1}(\omega),\omega^k),\,\,\,\,u'_k(\omega)=\RR'(u_{k-1}(\omega),u'_{k-1}(\omega),\omega^k),\nonumber\\
&\zeta_k(\omega)=\zeta(u_{k-1}(\omega),u'_{k-1}(\omega),\omega^k),\,\,\,\,\,\zeta'_k(\omega)=\zeta'(u_{k-1}(\omega),u'_{k-1}(\omega),\omega^k),\nonumber\\
&t_k(\omega)=t(\omega^k),\nonumber
\end{align} where $\omega=(\omega^1,\omega^2,...)\in\Omega$. Clearly, for  $U_k:=(u_k,u_k')$ assertion (i) of Proposition \ref{P:A} is satisfied.
 Since $\zeta_k,\zeta'_k$ and $t_k$ are sequences of independent random variables and $\{\zeta_k\}_{k=1}^\infty$ and $\{\zeta'_k\}_{k=1}^\infty$ are independent of $\{t_k\}_{k=1}^\infty$, the sequence $U_k$ is a Markov chain in the space $\HHH:=H\times H.$ \\
Let us introduce the stopping time
$$
\tau_d=\min\{k\ge0,\|u_k\|_1\vee\|u'_k\|_1\le d\}.
$$
\begin{lemma}\label{L:gn}
For any $d>0$ there are positive constants $\gamma$ and $C$ such
that
\begin{equation}\label{E:tau}
\E_U e^{\gamma\tau_d}\le C(1+\HH(u)+\HH(u'))\,\,\,\text{for all
$U:=(u,u')\in \HHH.$}
\end{equation}
\end{lemma}
\begin{proof}It is well known (e.g., see \cite{Has} or Proposition 2.3 in \cite{Shi04}) that inequality (\ref{E:tau}) will follow from  two statements below:
\begin{enumerate}
\item[(i)] There are positive constants $\delta, R$ and $C$ such that
\begin{equation}\label{E:taur}
\E_U e^{\delta\tau_R}\le C(1+\HH(u)+\HH(u'))\,\,\,\text{for all
$U\in \HHH$}.
\end{equation}
\item[(ii)] For any $R>0$ and $d>0$ there is an integer $l\ge1$ and a constant $p>0$ such that
\begin{equation}\label{E:taup}
\pP_U\{U_l\in\BBB_d\}\ge p\,\,\,\,\text{for any $U\in\BBB_R$},
\end{equation}where $\BBB_d=\{(u,u')\in\HHH:\|u\|_1\vee\|u'\|_1\le d\}$.
\end{enumerate}
The proof of (i) is similar to that of Lemma \ref{L:(i)}. To prove
(ii), we use the definition of $U_k=(u_k,u'_k)$, Lemma \ref{L:2} and
the Markov property:
$$
\pP_U\{\HH(u_l)+\HH(u_l')\le\big(\gamma^l(\HH(u)+\HH(u'))\big)\vee\big(d^2
\alpha\big)\}\ge\e^l,
$$for all $l\ge 1$, where $\e$ depends only on $d$. Choosing $l$ so large that $\gamma^l C_R<d^2\alpha$, where $C_R=\sup_{U\in\BBB_{R}}(\HH(u)+\HH(u'))$, we obtain (\ref{E:taup}).
\end{proof}
The proof of the following lemma is similar to that of Lemma
\ref{L:T}, and we shall not dwell on it.
\begin{lemma}\label{L:TT}
\begin{enumerate}
\item[(i)] There is a constant $M>0$ such that for any $U_0=(u_0,u_0')\in \HHH$ and an appropriate random integer $T=T(u_0,u_0')\ge1$ the following inequalities hold
\begin{align}
&\langle\|u_k\|_1^6+\|u_k'\|_1^6\rangle_0^n\le M \,\,\,\,\,\,\,\,\text{for  $n\ge T$,}\label{E:TT1}\\
&\E
T^p<\infty\,\,\,\,\,\,\,\,\,\,\,\,\,\,\,\,\,\,\,\,\,\,\,\,\,\,\,\,\,\,\,\,\,\,\,\,\,\,\,\,\,\,\,\text{for
all $p\ge1$.}\label{E:TT2}
\end{align}
\item[(ii)] For any $\delta\in(0,1)$ and $d>0$, there is a constant $R=R(\delta, d)>0$ such that
\begin{align}
\pP\{\langle\|u_k\|_1^6+\|u_k'\|_1^6\rangle_0^n\le R,\forall
n\ge0\}\ge\delta,\label{E:TT3}
\end{align}for any  $U_0=(u_0,u_0')\in \BBB_d$.
\end{enumerate}
\end{lemma}
For any $M>0$, we introduce the stopping times
\begin{align}
T_1(M)&=\min\{k\ge1:\langle\|u_i\|_1^6+\|u_i'\|_1^6\rangle_0^k>M\},\nonumber\\
T_2(M)&=\min\{k\ge1:\frac{1}{2}\langle\log t_i\rangle_0^k>M\},\nonumber\\
T_3(M)&=\min\{k\ge1:P_N u_k\neq P_N u'_k\},\nonumber\\
\sigma(M)&=T_1(M)\wedge T_2(M)\wedge T_3(M).\nonumber
\end{align}
\begin{lemma}\label{L:sigma}
For any $B\ge0,$ there is an integer $N'\ge1$ and a constant $M>0$
such that, if (\ref{E:HN*}) and  (\ref{E:HN**}) hold for some
integer $N\ge1$, then
\begin{align}
&\pP_U\{\sigma(M)=\infty\}\ge\frac{1}{2},\label{E:si1}\\
&\E_U[I_{\{\sigma(M)<\infty\}}\sigma(M)^p]<+\infty\,\,\,\,\,\,\,\text{for
all $p\ge1,$}\label{E:si2}
\end{align} where $U\in\BBB_d$, $d=\frac{1}{2C_N}$ and $C_N\ge 1$ is the constant in (\ref{E:kar}).
\end{lemma}
\begin{proof}
Let $M>0$ be sufficiently large and let $m\ge1.$ Then
\begin{align}\label{E:1*}
\{\sigma(M)=m\}\subset\{T_1(M)=m\}\cup\{T_2(M)=m\}\cup A_m,
\end{align}where $A_m=\{T_3(M)=m,T_1(M)\ge m,T_2(M)\ge m\}$. Note that
$$
A_m=\{P_Nu_m\neq P_Nu'_m,\sigma(M)>m-1\}.
$$It follows from Lemma \ref{L:P} that for $\pP_U$-a.e. $\omega\in\{\sigma(M)>m-1\}$, we have
$$
\|S_{t_m}(u_{m-1})-S_{t_m}(u'_{m-1})\|_1\le
2d(C\alpha_{N'+1}^{-\frac{1}{2}})^me^{2CMm}.
$$Choosing $N'$ so large that $\log\alpha_{N'+1}\ge2(2CM+\log C+2),$ we see that
$$
\|S_{t_m}(u_{m-1})-S_{t_m}(u'_{m-1})\|_1\le 2de^{-2m}.
$$Using Remark \ref{R:2}, construction of the space $(\Omega,\FF,\pP)$ and the Markov property, we obtain
\begin{equation}\label{E:TTT}
\pP_U(A_m)\le2dC_Ne^{-2m}=e^{-2m}.
\end{equation}
Let $T_2'$ be the random integer constructed in Lemma \ref{L:Ap} for
the sequence $\frac{1}{2}\log t_i$, and $T_1'$ be the random integer
in Lemma \ref{L:TT}. Then, it follows from the definition of $T_1$
and $T_2$ that
\begin{align}
&T_1I_{\{T_1<\infty\}}<T_1',\nonumber\\
&T_2I_{\{T_2<\infty\}}<T_2'\nonumber.
\end{align}To prove (\ref{E:si2}), note that
\begin{align}
\E_U[I_{\{\sigma<\infty\}}\sigma^p]&=\sum_{m=1}^\infty\pP\{\sigma(M)=m\}m^p\nonumber\\&\le\sum_{m=1}^\infty(\pP\{T_1'>m\}+\pP\{T_2'>m\}+\pP\{A_m\})m^p\nonumber\\&\le
C\sum_{m=1}^\infty(m^{-p-2}+e^{-2m})m^p<\infty,\nonumber
\end{align}where we used (\ref{E:1*}), (\ref{E:TTT}), (\ref{E:TT2}) and (\ref{E:Ap2}). \\To prove (\ref{E:si1}), we use (\ref{E:1*}) and (\ref{E:TTT}):
\begin{align}
\pP\{\sigma<\infty\}\le\pP\{T_1<\infty\}+\pP\{T_2<\infty\}+\frac{1}{e^2-1}.
\end{align} It follows from (\ref{E:TT3}) and (\ref{E:Mkgn}) that for any $\delta\in(0,1)$ there is $M=M(\delta,d)>0$ such that
$$
\pP\{T_1(M)<\infty\}+\pP\{T_2(M)<\infty\}<\delta.
$$Choosing $\delta=\frac{1}{2}-\frac{1}{e^2-1}$, we arrive at (\ref{E:si1}).
\end{proof}

To construct the random integer $\ell$ in Proposition \ref{P:A}, we
follow the ideas of \cite{Shi04}. Suppose that $N\ge1,$ $M$ and
$d\le 1$ are the constants in Lemma \ref{L:sigma}. Let $\rho_0$ be
the first hitting time of the set $\BBB_d$. If for some
$\omega\in\Omega$ we have
\begin{equation}\label{E:2*}
P_Nu_k=P_Nu_k',\,\,\,\langle\|u_i\|_1^6+\|u_i'\|_1^6\rangle_{\rho_0}^k\le
M,\,\,\,\frac{1}{2}\langle\log t_i\rangle_{\rho_0}^k\le
M\,\,\,\,\text{for all $k\ge\rho_0+1$}.
\end{equation}  we set $\ell(\omega)=\rho_0(\omega)$, otherwise, let $\rho_1'$ be the first time when one of the  conditions in (\ref{E:2*}) is not satisfied and let $\rho_1$
 be the first hitting time of the ball $\BBB_d$ after $\rho_1'$. Suppose that $\rho_1<\infty$ and (\ref{E:2*}) is verified for $\omega\in\Omega$, with $\rho_0$ replaced by $\rho_1$, then we set $\ell(\omega)=\rho_1(\omega)$.
  Continuing this process and using the same arguments as in \cite{Shi04}, one can
  show that $\ell$ is well defined for a.e. $\omega\in\Omega$ and satisfies (\ref{E:P5}). The other assertions of Proposition \ref{P:A} follow immediately from the construction.
\section{Appendix}
\subsection{Proof of inequality (\ref{E:gn1})}
 Let $u_0\in H$. Setting $u(t)=S_t(u_0)$, we have
\begin{equation}\label{E:DH}
\frac{\dd}{\dd t}\HH(u(t))=(-2\alpha\Delta u+\beta |u|^2u,\dot u),
\end{equation} where $(u,v)=\Re\int_D u\bar v\dd x.$ Since $u$ is the solution of (\ref{E:eq})-(\ref{E:ic}) with $\eta\equiv 0,$ we deduce from (\ref{E:DH}) that
\begin{align}\label{E:DH1}
\frac{\dd}{\dd t}\HH(u)&=(-2\alpha\Delta u+\beta |u|^2u,\nu\Delta
u-i\beta |u|^2u)\nonumber\\&\le-2\alpha\nu\|\Delta u\|^2+(\beta
|u|^2u,\nu\Delta u)+(2\alpha \Delta u,i\beta|u|^2u).
\end{align} It is clear that
\begin{align}\label{E:hh}
(2\alpha \Delta u,i\beta|u|^2u)\le2\alpha\beta(|\nabla u|^2,|u|^2),
\end{align}
\begin{align}\label{E:hh2}
(\beta|u|^2u,\nu\Delta u)=-\beta\nu(|u|^2,|\nabla
u|^2)-\beta\nu(u\nabla (|u|^2),\nabla u).
\end{align}Substituting (\ref{E:hh}) and (\ref{E:hh2}) into (\ref{E:DH1}) and noting that
$$
(u\nabla (|u|^2),\nabla u)=\Re\int_D u\nabla\bar u \nabla(|u|^2)\dd
x=\frac{1}{2}\Re\int_D(\nabla(|u|^2))^2\dd x\ge0,
$$we obtain
$$
\frac{\dd}{\dd t}\HH(u)\le-2\alpha\nu\|\Delta
u\|^2-\beta\nu(|u|^2,|\nabla u|^2)+2\alpha\beta(|u|^2,|\nabla u|^2).
$$Choosing $\alpha$ sufficiently small and applying Poincar\'e's inequality to the function $|u|^2$, we arrive at
\begin{equation}\label{E:hh3}
\frac{\dd}{\dd t}\HH(u)+\alpha\nu\|\Delta u\|^2\le-a \HH(u),
\end{equation} for some positive constant $a$. Application of Gronwall's inequality results in  (\ref{E:gn1}). Finally, note that the  integration of (\ref{E:hh3}) gives
\begin{equation}\label{E:hh4}
\alpha\nu\int_0^t\|\Delta u\|^2\dd s\le\HH(u_0).
\end{equation}
\subsection{Proof of inequalities (\ref{E:n1}) and (\ref{E:gn3})}
\textbf{Step 1.} Let $u(t)=S_t(u_0)$ and $u_0\in H.$ Then
\begin{equation}\label{E:l1}
u\cdot t^{\frac{1}{2}}\in C([0,\infty),H^2(D)).
\end{equation}Indeed, formally taking the scalar product of $-\Delta\dot u t$ and Equation (\ref{E:eq}) with $\eta\equiv 0,$ we obtain
$$
(\dot u-\nu\Delta u+i\beta|u|^2u,-\Delta\dot u t)=0.
$$ Integration of this equality in $t$ results in
\begin{align}\label{E:l2}
\frac{\nu}{2}t \|\Delta u\|^2+\int_0^t s\|\nabla \dot u\|^2\dd s\le
\frac{\nu}{2}\int_0^t\|\Delta u\|^2\dd s+\beta\int_0^t
s|(|u|^2u,\Delta\dot u)|\dd s.
\end{align}Note that
\begin{align}
\beta\int_0^t s|(|u|^2u,\Delta\dot u)|\dd s&\le\frac{1}{2}\int_0^t
s\|\nabla \dot u\|^2\dd s+C\int_0^t s\|\nabla
u\|_{L^4}^2\||u|^2\|_{L^4}^2\dd s\nonumber\\& \le\frac{1}{2}\int_0^t
s\|\nabla \dot u\|^2\dd s+C\int_0^t s\|\Delta u\|^2\|u\|_{L^8}^4\dd
s,
\end{align}
where we used Sobolev embedding $H^1(D)\hookrightarrow L^4(D)$.
Substituting this inequality into (\ref{E:l2}) and using Gronwall's
inequality, we arrive at
\begin{equation}
t\|\Delta u\|^2\le\int_0^t\|\Delta u\|^2\dd
s\cdot\exp\big(C\int_0^t\|u\|_{L ^8}^4\dd s\big).\label{E:VV}
\end{equation} Using the Gagliardo--Nirenberg inequality
\begin{equation}\label{E:GN}
\|u\|_{L^8}\le C \|u\|_{L^4}^{\frac{1}{2}}\|\Delta
u\|^{\frac{1}{2}},
\end{equation} and inequalities (\ref{E:gn1}) and (\ref{E:hh4}), we see that
\begin{align}
\exp(C\int_0^t\|u\|_{L^8}^4\dd s)&\le
\exp(C\int_0^t\|u\|_{L^4}^2\|\Delta u\|^2\dd s)\nonumber\\&\le\exp
(C \HH(u_0)^{\frac{3}{2}})\le C\exp(C\|u_0\|_1^6).\label{E:**}
\end{align}Now substituting (\ref{E:**}) into the right-hand side of (\ref{E:VV}), and using (\ref{E:hh4}), we obtain
\begin{equation}\label{E:l3}
\sup_{\tau\in[0,t]}\tau \|\Delta u (\tau)\|^2\le
C\exp(C\|u_0\|_1^6).
\end{equation}To prove (\ref{E:l1}), we use Galerkin's method, choosing as a base in $L^2(D)$ the set of normalized eigenfunctions of the Dirichlet Laplacian. It is easy to verify that (\ref{E:l3}) holds for Galerkin approximations. Then passing to the limit, we arrive at (\ref{E:l1}) and (\ref{E:l3}).
\\\textbf{Step 2.} Let $u_0,v_0\in H$ and $u=S_t(u_0)$, $v=S_t(v_0)$. Then we have the following estimate for $w=u-v:$
\begin{equation}\label{E:al}
\|\nabla w\|^2+\nu\int_0^t\|\Delta w\|^2\dd s\le C \|\nabla
w_0\|^2\exp(C(\|u_0\|_1^6+\|v_0\|_1^6)).
\end{equation}where $w_0=u_0-v_0$ and $C$ is a positive constant.
Indeed, $w$ is a solution of the following equation
 \begin{equation}\label{E:al1}
\dot w-\nu\Delta w+i\beta(|u|^2u-|v|^2v)=0.
\end{equation}Taking the scalar product of this equation with $-\Delta w$ and  integrating the resulting equality in $t$, we see that
\begin{align}\label{E:a2}
&\frac{1}{2}\|\nabla w\|^2+\nu\int_0^t\|\Delta w\|^2\dd s\le\frac{1}
{2}\|\nabla w_0\|^2+\beta\int_0^t|(|u|^2u-|v|^2v,\Delta w)|\dd
s\nonumber\\&\le\frac{1}{2}\|\nabla
w_0\|^2+\frac{\nu}{2}\int_0^t\|\Delta w\|^2\dd
s+C\int_0^t\||u|^2+|v|^2\|_{L^4}^2\|\nabla w\|^2\dd s.
\end{align} We deduce from Gronwall's inequality:
$$
\|\nabla w\|^2\le\|\nabla
w_0\|^2\exp(C\int_0^t\||u|^2+|v|^2\|_{L^4}^2\dd s).
$$Now substituting this inequality into the right-hand side of (\ref{E:a2}) and using (\ref{E:**}), we arrive at (\ref{E:al}).
\\\textbf{Step 3.} Taking the scalar product of (\ref{E:al1}) with $-t\Delta\dot w$ and integrating the resulting equality, we obtain
\begin{align}\label{E:a3}
&\frac{\nu}{2}t\|\Delta w\|^2+\int_0^ts\|\nabla\dot w\|^2\dd
s\le\frac{\nu}{2}\int_0^t\|\Delta w\|^2\dd
s\nonumber\\&+\frac{\beta^2}{2}\int_0^ts\|\nabla(|u|^2u-|v|^2v)\|^2\dd
s+\frac{1}{2}\int_0^t s\|\nabla\dot w\|^2\dd s.
\end{align}Using H\"older's inequality, we see that
\begin{align}\label{E:a4}
&\int_0^ts\|\nabla(|u|^2u-|v|^2v)\|^2\dd s\nonumber\\&\le
C\int_0^ts(\||u|^2\nabla w\|^2+\|w u\nabla v\|^2+\|wv\nabla
v\|^2)\dd s\nonumber\\&\le C\int_0^t (s\|\Delta
w\|^2\|u\|_{L^8}^4+s\|w\|_{L^8}^2\|\nabla
v\|^2_{L^4}(\|v\|_{L^8}^2+\|u\|_{L^8}^2))\dd s.
\end{align}Substituting (\ref{E:a4}) into (\ref{E:a3}) and using Gronwall's inequality, we arrive at
\begin{align}\label{E:a5}
t\|\Delta w\|^2&\le C\Big[\int_0^t\|\Delta w\|^2\dd
s+\int_0^ts\|w\|_{L^8}^2\|\nabla
v\|^2_{L^4}(\|v\|_{L^8}^2+\|u\|_{L^8}^2)\dd
s\Big]\nonumber\\&\times\exp(C\int_0^t \|u\|_{L^8}^4\dd s).
\end{align} By the Cauchy--Schwarz inequality,
\begin{align}\label{E:pp1}
P:=&\int_0^ts\|w\|_{L^8}^2\|\nabla
v\|^2_{L^4}(\|u\|_{L^8}^2+\|v\|_{L^8}^2)\dd s\le
C\sup_{[0,t]}s\|\Delta
v\|^2\nonumber\\&\times\Big(\int_0^t\|w\|_{L^8}^4\dd
s\Big)^{\frac{1}{2}}\Big[\Big(\int_0^t\|u\|_{L^8}^4\dd
s\Big)^{\frac{1}{2}}+\Big(\int_0^t\|v\|_{L^8}^4\dd
s\Big)^{\frac{1}{2}}\Big].
\end{align} To estimate the right-hand side of this inequality, note that
\begin{align}\label{E:LLL}
\sup_{[0,t]}s\|\Delta v\|^2&\Big[\Big(\int_0^t\|u\|_{L^8}^4\dd
s\Big)^{\frac{1}{2}}+\Big(\int_0^t\|v\|_{L^8}^4\dd
s\Big)^{\frac{1}{2}}\Big]\nonumber\\&\le
C\exp(C(\|u_0\|_1^6+\|v_0\|_1^6)),
\end{align} where we used (\ref{E:**}) and (\ref{E:l3}). Using (\ref{E:GN}) and (\ref{E:al}), we see that
\begin{align}
\Big(\int_0^t\|w\|_{L^8}^4\dd s\Big)^{\frac{1}{2}}&\le
C\sup_{[0,t]}\|\nabla w\|\Big(\int_0^t\|\Delta w\|^2\dd
s\Big)^{\frac{1}{2}}\nonumber\\&\le C \|\nabla w_0\|^2
\exp(C(\|u_0\|_1^6+\|v_0\|_1^6)).\label{E:MMM}
\end{align} We deduce from (\ref{E:LLL}) and (\ref{E:MMM}) that
\begin{align}
P\le C \|\nabla
w_0\|^2\exp(C(\|u_0\|_1^6+\|v_0\|_1^6)).\label{E:JJJ}
\end{align}Finally, substituting (\ref{E:JJJ}) into (\ref{E:a5}), and using (\ref{E:al}) and (\ref{E:**}), we arrive at (\ref{E:gn3}).

\subsection{Proof of Lemma \ref{L:MP}}

Let $\NN_t$ be defined by (\ref{E:Nt}). Then, for $0\le s\le t$,
$\NN_t-\NN_s$ is a Poisson random variable with parameter
$\lambda(t-s),$ independent of $\FF_t$ (e.g., see \cite{Kar}), where
$\FF_t$ is  defined in Section \ref{S:MC}. Let $\zeta$ be defined by
(\ref{E:zeta}), then we have
\begin{equation}\label{E:11}
\zeta(t)-\zeta(s)=\sum_{k=\NN_s+1}^{\NN_t}\eta_k.
\end{equation} It follows from (\ref{E:11}) that $\zeta$ has independent increments. Using  (\ref{E:11}) and the fact that the distribution of $\NN_t-\NN_s$ depends only on $t-s$, it is easy to see that the distributions of processes $\zeta(\cdot)$ and $\zeta(\cdot+s)-\zeta(s)$ coincide:
\begin{equation}\label{E:12}
\DD(\zeta(t),t\ge0)=\DD(\zeta(t+s)-\zeta(s),t\ge0),
\end{equation} for any $s\ge0$. Note that $u(t)$ is determined by $\{\zeta(\tau):0\le\tau\le t
\}$ and $u(t)$ is $\FF_t$-measurable. We have
\begin{align}
\pP\{u(t,u_0,\{\zeta(\tau)&:0\le\tau\le t
\})\in\Gamma|\FF_s\}\nonumber\\&=\pP\{u(t-s,u_s,\{\zeta(\tau)-\zeta(s):s\le\tau\le
t\})\in\Gamma|\FF_s\}\nonumber\\&=\pP\{u(t-s,v,\{\zeta(\tau)-\zeta(s):s\le\tau\le
t \})\in\Gamma\}\big|_{v=u_s},
\end{align}for any $\Gamma\in\BB(H)$, where we used the independence of increments of $\zeta.$ Using (\ref{E:12}), we arrive at
\begin{equation}
\pP\{u(t,u_0)\in\Gamma|\FF_s\}=\pP\{u(t-s,v)\in\Gamma\}\big|_{v=u_s},
\end{equation} which completes the proof of the lemma.
\subsection{Proof of Lemma \ref{L:T}}
Let us introduce the random variable
$$
T=\min\{n\ge1:\,\,\,\frac{1}{k}|M_k|\le1\,\,\,\,\text{for all $k\ge
n$}\},
$$ where $\min{\{\emptyset\}}=+\infty$.
 It is easy to see that $\pP\{T=\infty\}=0$, as
$$
\pP\{T=\infty\}\le\sum_{k=m}^\infty\pP\{\frac{1}{k}|M_k|>1\}\le
C\sum_{k=m}^\infty\frac{1}{k^2}\rightarrow0,\,\,\,\,m\rightarrow\infty,
$$ where we used (\ref{E:Mk}) with $p=2$ and Chebyshev's inequality. To estimate the moments of $T$, we use (\ref{E:Mk}) with $l=p+2$:
\begin{eqnarray}
\E T^p=\sum_{n=1}^\infty\pP\{T=n\}n^p&\le&
1+\sum_{k=1}^\infty\pP\{\frac{1}{k}|M_k|>1\}(k+1)^p\nonumber\\&\le&1+C\sum_{k=1}^\infty
k^{-l}(k+1)^p<+\infty.\nonumber
\end{eqnarray}To prove (\ref{E:Mkgn}), we  use  (\ref{E:Mk}) with $p=2$ and Chebyshev's inequality
\begin{align}
\pP\{\frac{|M_k|}{k}\le R,\forall
k\ge1\}\ge1-\sum_{k=1}^\infty\pP\{\frac{|M_k|}{k}>
R\}\ge1-\frac{C_2}{R^4}\sum_{k=1}^\infty\frac{1}{k^2}.\nonumber
\end{align}Choosing $R$ sufficiently large, we obtain (\ref{E:Mkgn}).
\subsection{Proof of Theorem  \ref{T:BK}}
Let $u_t$ be the trajectory of (\ref{E:eq})-(\ref{E:pr}) with
$u_0\equiv0$. It suffices to show that the family $\DD(u_t)$ is
tight in $H$. Let $\NN_t$ be defined by (\ref{E:Nt}). First we shall
show that the family $\DD(u_{\NN_t})$ is tight. By Ulam's theorem,
there is a compact $K_\e^1$ such that $\pP\{\eta_1\notin
K_\e^1\}\le\frac{\e}{2}$. Using the independence of $\{\eta_p\}$ and
$\NN_t$, we obtain
\begin{align}
\pP\{\eta_{\NN_t}\notin
K_\e^1,\NN_t\neq0\}&=\sum_{p=1}^\infty\pP\{\eta_p\notin K_\e^1,
\NN_t=p\}\nonumber\\&=\sum_{p=1}^\infty\pP\{\eta_p\notin
K_\e^1\}\{\NN_t=p\}\le\frac{\e}{2}.\nonumber
\end{align}Using (\ref{E:gnk}), it is easy to show that there is a
constant $M>0$ such that
\begin{equation}\label{E:bav0}
\E\big(\|u_{\tau_{\NN_t-1}}\|_1 I_{\{\NN_t\neq0\}}\big)\le
M,\,\,\,\,\,\,\,\,\,\,\,\, \text{for all $t\ge0.$}\nonumber
\end{equation} Let $R_\e\ge\frac{4M}{\e}$. By the Chebyshev
inequality, we have
\begin{equation}\label{E:bav4}
\pP\{\|u_{\tau_{\NN_t-1}}\|_1 I_{\{\NN_t\neq0\}}\ge R_\e\}\le
\frac{M}{R_\e}\le\frac{\e}{4}.
\end{equation} Define $B_\e=\{v\in H: \|v\|_1\le R_\e\}$ and
$K_\e^2=S_{[a,b]}(B_{R_\e})$, where $b>0, a>0$. Then $K_\e^2$ is
compact in $H$. We deduce from (\ref{E:bav4}) that
\begin{align}
&\pP\{S_{t_{\NN_t}}(u_{\tau_{\NN_t-1}})\notin K_\e^2,
\NN_t\neq0\}\le\pP\{t_{\NN_t}>b, \NN_t\neq0\}+\pP\{t_{\NN_t}<a,
\NN_t\neq0\}\nonumber\\&+\pP\{u_{\tau_{\NN_t-1}}\notin B_\e,
\NN_t\neq0\}\le\sum_{p=1}^\infty(\pP\{t_p>b, \NN_t=p\}+\pP\{t_p<a,
\NN_t=p\})+\frac{\e}{4}\nonumber\\&\le\pP\{t_1>b\}+\pP\{t_1<a\}+\frac{\e}{4}\le\frac{\e}{2},\nonumber
\end{align}if $b$ is sufficiently large and $a$ is sufficiently small. Let $K_\e=K_\e^1+K_\e^2$. We can assume that $0\in K_\e$. As $u_{\NN_t}=0$, if $\NN_t=0,$ we
have
\begin{align}
\pP\{u_{\tau_{\NN_t}}\notin K_\e\}&=\pP\{u_{\tau_{\NN_t}}\notin
K_\e,\NN_t=0\}+\pP\{u_{\tau_{\NN_t}}\notin
K_\e,\NN_t\neq0\}\nonumber\\&\le \pP\{\eta_{\NN_t}\notin
K_\e^1,\NN_t\neq0\} +\pP\{S_{t_{\NN_t}}(u_{\tau_{\NN_t-1}})\notin
K_\e^2, \NN_t\neq0\}\le\e.\nonumber
\end{align} Thus $\DD(u_{\tau_{\NN_t}})$ is tight in $H$. \\
Define $T_\e=S_{[0,\infty)}(K_\e)$ and note that
$u_t=S_{t-\tau_{\NN_t}}(u_{\tau_{\NN_t}})$, $t\ge0$. Then $T_\e$ is
compact in $H$. Finally, we have
$$
\pP\{u_t\notin T_\e\}\le\pP\{u_{\tau_{\NN_t}}\notin K\}\le\e.
$$The proof of the other assertion of the theorem is standard.

\end{document}